\documentclass[12pt]{article}
\usepackage{enumerate}
\usepackage{amsfonts}
\usepackage{amssymb}
\usepackage{amsmath}
\usepackage{graphicx}
\usepackage{epsfig}

\begin{document}

\title{Gaussian classical-quantum channels: \\ gain of entanglement-assistance}
\author{A. S. Holevo \\
Steklov Mathematical Institute, Moscow}
\date{}
\maketitle

\begin{abstract}
In the present paper we introduce and study Bosonic Gaussian
classical-quantum (c-q) channels; the embedding of the classical input into
quantum is always possible and therefore the classical entanglement-assisted capacity $C_{ea}$ under appropriate input
constraint is well defined. We prove a general property of entropy increase
for weak complementary channel, that implies the equality $C=C_{ea}$ (where $C$ is the unassisted capacity) for
certain class of c-q Gaussian channel under appropriate energy-type
constraint. On the other hand, we show by explicit example that the
inequality $C<C_{ea}$ is not unusual for constrained c-q Gaussian channel.
\end{abstract}

\section{Introduction}

In finite dimension a classical-quantum or quantum-classical channel can
always be represented as a quantum channel, by embedding the classical input
or output into quantum system. Then it makes sense to speak about
entanglement-assisted capacity $C_{ea}$ \cite{BSST}, \cite{E} of such a
channel, in particular, to compare it with the unentangled classical
capacity $C$. An interesting observation in \cite{BSST} was that
entanglement-assisted communication may be advantageous even for \emph{
entanglement-breaking} channels such as depolarizing channel with
sufficiently high error probability. In the paper \cite{H2} we considered the case of
quantum-classical (measurement) channels, showing that generically $C<C_{ea}$
for such channels. For infinite dimensional (in particular, continuous
variable) systems an embedding of the classical output into quantum is not
always possible, however entanglement-assisted transmission still makes
sense \cite{H2}; in particular this is the case for Bosonic Gaussian q-c
channels. The measurement channels demonstrate the gain of entanglement
assistance in the most spectacular way.

On the contrary, as shown in \cite{Sh-19}, finite dimensional c-q channels
(preparations) are \emph{essentially} characterized by the property of
having no gain of entanglement assistance, in this sense being ``more
classical'' than measurements. In the present paper we study Bosonic
Gaussian c-q channels; we observe that the embedding of the classical input
into quantum is always possible and $ C_{ea}$ under the input constraint is
thus well defined. We prove a general property of entropy increase for the
weak complementary channel, that implies equality $C=C_{ea}$ for certain
class of c-q Gaussian channel under appropriate energy-type constraint. On
the other hand, we show by explicit example that the inequality $C<C_{ea}$
is not unusual for \emph{constrained} c-q Gaussian channels.

\section{Bosonic Gaussian Systems}

The main applications of infinite-dimensional quantum information theory are
related to Bosonic systems, for detailed description of which we refer to
Ch. 12 in \cite{H-SSQT}. Let $\mathcal{H}_{A}$ be the representation space
of the Canonical Commutation Relations (CCR)
\begin{equation}
W(z_{A})W(z_{A}^{\prime })=\exp \left( -\frac{i}{2}z_{A}^{t}\Delta
_{A}z_{A}^{\prime }\right) W(z_{A}^{\prime }+z_{A})  \label{CCR}
\end{equation}%
with a coordinate symplectic space $(Z_{A},\Delta _{A})$ and the Weyl system
$W_{A}(z)=\exp (iR_{A}\cdot z_{A});\,z_{A}\in Z_{A}$. Here $R_{A}$ is the
row-vector of the canonical variables in $\mathcal{H}_{A}$, and $\Delta _{A}$
is the canonical skew-symmetric commutation matrix of the components of $%
R_{A}$,
\begin{equation}
\Delta =\mathrm{diag}\left[
\begin{array}{cc}
0 & 1 \\
-1 & 0%
\end{array}%
\right] _{j=1,\dots ,s}.  \label{cf}
\end{equation}

Let $(Z_{A},\Delta _{A}),(Z_{B},\Delta _{B})$ be the symplectic spaces of
dimensions $2s_{A},2s_{B},$ which will describe the input and the
output of the channel (here $\Delta _{A},\Delta _{B}$ have the canonical
form (\ref{cf})), and let $W_{A}(z_{A}),W_{B}(z_{B})$ be the Weyl operators
in the Hilbert spaces $\mathcal{H}_{A},\mathcal{H}_{B}$ of the corresponding
Bosonic systems. A centered Gaussian channel $\Phi :\mathfrak{T}(\mathcal{H}%
_{A})\rightarrow \mathfrak{T}(\mathcal{H}_{B})$ is defined via the action of
its dual $\Phi ^{\ast }$ on the Weyl operators:
\begin{equation}
\Phi ^{\ast }[W_{B}(z_{B})]=W(Kz_{B})\exp \left[ -\frac{1}{2}z_{B}^{t}\alpha
z_{B}\right] ,  \label{gaus-ch}
\end{equation}%
where $K$ is matrix of a linear operator $Z_{B}\rightarrow Z_{A}$, and $\alpha $ is real
symmetric matrix satisfying
\begin{equation}
\alpha \geq \pm \frac{i}{2}\left( \Delta _{B}-K^{t}\Delta _{A}K\right),
\label{nid}
\end{equation}
where $\Delta _{B}-K^{t}\Delta _{A}K \equiv \Delta _{K}$ is a real
skew-symmetric matrix.

We will make use of the unitary dilation of the channel $\Phi $ constructed
in \cite{cegh1} (see also \cite{H-SSQT}). Consider the composite Bosonic
system $AD=BE$ with the Hilbert space $\mathcal{H}_{A}\otimes \mathcal{H}%
_{D}\simeq \mathcal{H}_{B}\otimes \mathcal{\ H}_{E}$ corresponding to the
symplectic space $Z=Z_{A}\oplus Z_{D}=Z_{B}\oplus Z_{E},$ where $%
(Z_{E},\Delta _{E})\simeq (Z_{A},\Delta _{A})$. Thus $[R_{A}\,R_{D}]=[R_{B}%
\,R_{E}]$ describe two different splits of the set of canonical observables
for the composite system. Here $A$ and $B$ refer to input and output,
while $D$ and $E$ to input and output environments.
The channel $\Phi $ is then described by the
linear input-output relation (preserving the commutators)
\begin{equation}
R_{B}^{\prime }=R_{A}K+R_{D}K_{D},  \label{ior}
\end{equation}%
where the system $D$ is in a centered Gaussian state $\rho _{D}$ with the
covariance matrix $\alpha _{D}$ such that
\begin{equation*}
\alpha =K_{D}^t\alpha _{D}K_{D}.
\end{equation*}%
(for simplicity of notations we write $R_{A},\dots $ instead of $%
R_{A}\otimes I_{D},\dots $). It is shown that the commutator-preserving
relation (\ref{ior}) can be complemented to the full linear canonical
transformation by putting
\begin{equation}
R_{E}^{\prime }=R_{A}L+R_{D}L_{D},  \label{iocomp}
\end{equation}%
where $\left( 2s_{A}\right) \times \left( 2s_{E}\right) -$ matrix $L$ and $%
\left( 2s_{D}\right) \times \left( 2s_{A}\right) -$ matrix $L_{D}$ are such
that the square $2\left( s_{A}+s_{D}\right) \times 2\left(
s_{B}+s_{E}\right) -$ matrix
\begin{equation}
T=\left[
\begin{array}{cc}
K & L \\
K_{D} & L_{D}%
\end{array}%
\right]  \label{bltr}
\end{equation}%
is symplectic, i.e. satisfies the relation
\begin{equation*}
T^{t }\left[
\begin{array}{cc}
\Delta _{A} & 0 \\
0 & \Delta _{D}%
\end{array}%
\right] T=\left[
\begin{array}{cc}
\Delta _{B} & 0 \\
0 & \Delta _{E}%
\end{array}%
\right] ,
\end{equation*}%
which is equivalent to
\begin{eqnarray}
\Delta _{B} &=&K^{t}\Delta _{A}K+K_{D}^{t}\Delta _{D}K_{D},\quad  \label{com}
\\
0 &=&K^{t}\Delta _{A}L+K_{D}^{t }\Delta _{D}L_{D},  \label{com1} \\
\Delta _{E} &=&L^{t}\Delta _{A}L+L_{D}^{t }\Delta _{D}L_{D}.  \label{com2}
\end{eqnarray}

Denote by the $U_{T}$ the unitary operator in $\mathcal{H}_{A}\otimes
\mathcal{H}_{D}\simeq \mathcal{H}_{B}\otimes \mathcal{H}_{E}$ implementing
the symplectic transformation $T$ so that
\begin{equation}
\lbrack R_{B}^{\prime }\,R_{E}^{\prime }]=U_{T}^{\ast
}[R_{B}\,R_{E}]U_{T}=[R_{A}\,R_{D}]T.  \label{deistvo}
\end{equation}%
Then we have the unitary dilation
\begin{equation}
\Phi ^{\ast }[W_{B}(z_{B})]=\mathrm{Tr}_{D}\left( I_{A}\otimes \rho
_{D}\right) U_{T}^{\ast }\left( W_{B}(z_{B})\otimes I_{E}\right) U_{T}.
\label{udi1}
\end{equation}%
The \emph{weakly complementary} channel \cite{cegh1} is then
\begin{equation*}
\left( \tilde{\Phi}^{w}\right) ^{\ast }[W_{E}(z_{E})]=\mathrm{Tr}_{D}\left(
I_{A}\otimes \rho _{D}\right) U_{T}^{\ast }\left( I_{B}\otimes
W_{E}(z_{E})\right) U_{T}.
\end{equation*}%
The equation (\ref{iocomp}) is nothing but the input-output relation for the
weakly complementary channel which thus acts as
\begin{equation}
\left( \tilde{\Phi}^{w}\right) ^{\ast }[W_{E}(z_{E})]=W_{A}(Lz_{E})\exp %
\left[ -\frac{1}{2}z_{E}^{t}L_{D}^{t}\alpha _{D}L_{D}z_{E}\right] .
\label{Gc}
\end{equation}

In the case of pure state $\rho _{D}=|\psi _{D}\rangle \langle \psi _{D}|$
the relation (\ref{udi1}) amounts to the Stinespring representation for the
channel $\Phi $ with the isometry $V=U_{T}|\psi _{D}\rangle ,$ implying that
$\tilde{\Phi}^{w}$ is the complementary channel $\tilde{\Phi}$ (see e.g.
\cite{H-SSQT}).

\section{A property of Gaussian classical-quantum channels}
\label{2}
Usually classical-quantum (c-q) channel is understood as a mapping $x\rightarrow \rho _{x}$ of the classical alphabet
$\mathcal{X}=\{x\}$ into density operators in a Hilbert space. In the case of continuous alphabet there is
no problem with embedding c-q channel into a quantum channel (as distinct from q-c channel, see \cite{H2}).
Intuitively, let $\mathcal{X}$ be a continual domain with measure $dx$, then the required embedding is
\begin{equation*}
\Phi [\rho ]=\int_{\mathcal{X}}\langle x|\rho |x\rangle \rho _{x}dx,
\end{equation*}
where $\left\{ |x\rangle ;x\in \mathcal{X}\right\} $ is a Dirac's system
satisfying $\langle x|x^{\prime }\rangle =$ $\delta (x-x^{\prime }).$ Here
$\Phi$ maps density operators into density operators. Notice that the
range of the dual channel $\Phi ^{\ast }$ consists of bounded operators diagonal
in the $x$-representation.

In general, we call a quantum channel $\Phi $ \emph{classical-quantum} (c-q) if the
range of $\Phi ^{\ast }$ consists of commuting operators. By using a structure
theorem for Abelian algebras of operators in a Hilbert space, it is then not
difficult to see that such a definition is essentially equivalent to the usual understanding.
It follows from (\ref{CCR}) that the necessary and sufficient condition for a Bosonic Gaussian
channel (\ref{gaus-ch}) to be c-q is
\begin{equation}
K^{t}\Delta _{A}K=0.  \label{cl-qu}
\end{equation}
Thus $\Delta _{K}=\Delta _{B}$ and therefore $\det \Delta _{K}\neq 0.$ Under
this condition it was shown in \cite{H3} that in the unitary dilation
described above one can take $s_{E}=s_{A},\,s_{D}=s_{B}$ (and in fact $%
E=A,D=B$). We call such a dilation ``minimal'' as it is indeed such at least
in the case of the pure state $\rho _{D}$, as follows from \cite{cegh1}. The
condition (\ref{nid}) then amounts to
\begin{equation}
\alpha \geq \pm \frac{i}{2}\Delta _{B},  \label{unc}
\end{equation}%
saying that $\alpha $ is a covariance matrix of a centered Gaussian state $%
\rho _{D}$. We say that the channel has \emph{minimal noise} if $\rho _{D}$
is a pure state, which is equivalent to the fact that $\alpha $ is a minimal
solution of the inequality (\ref{unc}). In quantum optics such channels are
called quantum-limited.

Let us explain how this notion of c-q channel agrees with the usual one
in the case of Bosonic Gaussian channels. The condition (%
\ref{cl-qu}) means that the components of the operator $R_{A}K$ all commute,
hence their joint spectral measure is a sharp observable, and their
probability distribution $\mu _{\rho }(d^{2n}z)$ can be arbitrarily sharply
peaked around any point $z=\mathsf{E}_{\rho }(R_{A}K)^{t}=K^{t}m$ in the
support $\mathcal{X}$ of this measure by appropriate choice of the state $%
\rho $. Here $\mathsf{E}_{\rho }$ denotes expectation with respect to $\rho $
and $m=\mathsf{E}_{\rho }(R_{A})^{t}$, hence $\mathcal{X}=\mathbf{Ran}%
K^{t}\subseteq Z_{B}$. Thus in this case it is natural to identify $\Phi $
as c-q channel determined by the family of states $z\rightarrow W(z)\rho
_{B}W(z)^{\ast };z\in \mathcal{X}$.

\textbf{Proposition 1.} \emph{Let }$\Phi $ \emph{be a Gaussian c-q channel,
then the weak complementary} $\tilde{\Phi}^{w}$ \emph{in the minimal unitary
dilation has nonnegative entropy gain:}
\begin{equation*}
S(\tilde{\Phi}^{w}[\rho ])-S(\rho )\geq 0\quad \text{\textrm{for all \ \ \ }}%
\rho .
\end{equation*}

\emph{In particular if} $\Phi $ \emph{has minimal noise, then this holds for
the complementary channel $\tilde{\Phi}$, implying}
\begin{equation}\label{IleS}
I(\rho ,\Phi )\leq S(\Phi \lbrack \rho ]),
\end{equation}%
\emph{where}
\begin{equation*}
I(\rho ,\Phi )=S(\rho )+S(\Phi \lbrack \rho ])-S(\tilde{\Phi}[\rho ])
\end{equation*}%
\emph{is the quantum mutual information.}

\emph{Proof.} Taking into account (\ref{cl-qu}), the relation (\ref{com})
becomes
\begin{equation}
\quad \Delta _{B}=K_{D}^{t}\Delta _{D}K_{D}.  \label{iskom2}
\end{equation}%
We consider the minimal dilation for which $\Delta _{D}=\Delta _{B}$, $%
\Delta _{E}=\Delta _{A}$, hence $K_{D}$ is a symplectic $2s_{B}\times 2s_{B}-
$ matrix. Then (\ref{com1}) implies
\begin{equation*}
L_{D}=-\left( K_{D}^{t}\Delta _{D}\right) ^{-1}K^{t}\Delta _{A}L.
\end{equation*}%
Substituting (\ref{com2}) gives $\Delta _{E}=L^{t}ML,$ where
\begin{eqnarray*}
M &=&\Delta _{A}+\Delta _{A}K\left( \Delta _{D}K_{D}\right) ^{-1}\Delta
_{D}\left( K_{D}^{t}\Delta _{D}\right) ^{-1}K^{t}\Delta _{A} \\
&=&\Delta _{A}+\Delta _{A}KK_{D}^{-1}\Delta _{D}^{-1}\left( K_{D}^{t}\right)
^{-1}K^{t}\Delta _{A} \\
&=&\Delta _{A}+\Delta _{A}K\Delta _{B}^{-1}K^{t}\Delta _{A}.
\end{eqnarray*}%
Therefore $1=\left( \det L\right) ^{2}\det M,$ where
\begin{eqnarray*}
\det M &=&\det \left( \Delta _{A}+\Delta _{A}K\Delta _{B}^{-1}K^{t}\Delta
_{A}\right) \\
&=&\det \left( I_{2s_{A}\times 2s_{A}}+K\Delta _{B}^{-1}K^{t}\Delta
_{A}\right) .
\end{eqnarray*}%
Due to (\ref{cl-qu}) the matrix $N=K\Delta _{B}^{-1}K^{t}\Delta _{A}$
satisfies $N^{2}=0,$ hence it has only zero eigenvalues. Therefore $%
I_{2s_{A}\times 2s_{A}}+N$ \ has only unit eigenvalues, implying $\det M=1$
and hence $\left\vert \det L\right\vert =1.$

By relation (\ref{Gc}), the channel $\tilde{\Phi}^{w}$ is the Gaussian
channel with the operator $L$ playing the role of $K.$ By using a result of
\cite{H1}, we have
\begin{equation*}
S(\tilde{\Phi}^{w}[\rho ])-S(\rho )\geq \log |\det L|=0.\qquad \square
\end{equation*}

\textbf{Proposition 2}. \emph{Let} $\Phi $ \emph{be a Gaussian c-q channel
with minimal noise } $\alpha $, \emph{such that } $\mathbf{Ran}K^{t}=Z_{B}$,
\emph{satisfying the input constraint\footnote{The trace here is understood in the sense of extended expectation, as in \cite{H1}.} }
\begin{equation}
\mathrm{Tr}\rho H\leq E,  \label{constraint}
\end{equation}%
\emph{where} $H=RK\epsilon K^{t}R^{t}$ \emph{and} $\epsilon $ \emph{is real
symmetric strictly positive definite matrix. }

\emph{Then denoting }$C(E)$ (\emph{resp.} $C_{ea}(E)$) \emph{the classical
(resp. entanglement-assisted) capacity of the channel under the constraint (%
\ref{constraint}), }
\begin{equation}
C(E)=C_{ea}(E)=\sup_{\rho :\mathrm{Tr}\rho H\leq E}S(\Phi \lbrack \rho ]).
\label{main}
\end{equation}

An important condition here is $\mathbf{Ran}K^{t}=Z_{B}$, as we shall see in
the next Section. The form of the operator $H=RK\epsilon K^{t}R^{t}$ is such
that the constraint is expressed only in terms of the input observables of
the c-q channel. Without it one could hardly expect the equality (\ref{main}), although this requires further investigation. On the other hand,
assumption of minimality of the noise seems to be related to the method of
the proof and probably could be relaxed, with the last expression in (\ref{main})
replaced by the supremum of $\chi$-function.

\bigskip \textbf{Lemma.} \emph{Under the assumption }(\ref{cl-qu})\emph{\
there exists a sequence of real symmetric} $\left( 2s_{A}\right) \times
\left( 2s_{A}\right) -$\emph{matrices} $\gamma _{n}$ \emph{satisfying the
conditions:}

\begin{enumerate}
\item $\gamma _{n}\geq \pm \frac{i}{2}\Delta _{A};$

\item $K^{t}\gamma _{n}K\rightarrow 0.$
\end{enumerate}

\emph{Proof.} The assumption (\ref{cl-qu}) means that the subspace $%
\mathcal{N}=\mathrm{Ran}K\subseteq Z_{A}$ is isotropic, i.e. such that $%
\Delta _{A}$ is degenerate on it. From the linear algebra it is known that
there is a symplectic basis in $Z_{A}$ of the form $\left\{ e_{1},\dots
,e_{k},h_{1},\dots ,h_{k},g_{1},\dots \right\} ,$ where $\left\{ e_{1},\dots
,e_{k}\right\} $ is a basis in  $\mathcal{N},\left\{ h_{1},\dots
,h_{k}\right\} $ span the isotropic subspace $\mathcal{N}^{\prime }$ and
are such that $e_{i}^{t}\Delta _{A}h_{j}=\delta _{ij},$ and $\left\{
g_{1},\dots \right\} $ span the symplectic orthogonal complement of $%
\mathcal{N}+\mathcal{N}^{\prime }.$ Then $\Delta _{A}$ has the block matrix
form in this basis%
\begin{equation*}
\Delta _{A}=\left[
\begin{array}{ccc}
0 & I_{k} & 0 \\
-I_{k} & 0 & 0 \\
0 & 0 & \Delta _{g}%
\end{array}%
\right] .
\end{equation*}%
Let $\varepsilon _{n}$ be a sequence of positive numbers converging to zero,
then
\begin{equation*}
\gamma _{n}=\left[
\begin{array}{ccc}
\varepsilon _{n}I_{k} & 0 & 0 \\
0 & \frac{1}{4\varepsilon _{n}}I_{k} & 0 \\
0 & 0 & \gamma _{g}%
\end{array}%
\right] ,
\end{equation*}%
where $\gamma _{g}\geq \pm \frac{i}{2}\Delta _{g},$ satisfies the condition
1, and $K^{t}\gamma _{n}K=\varepsilon _{n}K^{t}K\rightarrow 0.\quad \square $

\emph{Proof of Proposition 2.} According to the general version of the finite-dimensional
result of \cite{E} proven in \cite{H-Sh},
\begin{equation}
C_{ea}(E)=\sup_{\rho :\mathrm{Tr}\rho H\leq E}I(\rho, \Phi ).
\end{equation}
This version makes the only assumption that $H$ is positive self-adjoint operator,
allowing the constraint set to be non-compact, which is important for our considerations in Sec. \ref{3}.
Due to (\ref{IleS}), it is then sufficient to show that
\begin{equation*}
C(E)\geq \sup_{\rho :\mathrm{Tr}\rho H\leq E}S(\Phi \lbrack \rho ]).
\end{equation*}%
We first consider the supremum in the right-hand side. Since the constraint
operator $H=RK\epsilon K^{t}R^{t}$ is quadratic in the canonical variables $%
R,$ the supremum can be taken over (centered) Gaussian states. Since the
entropy of Gaussian state with covariance matrix $\alpha $ is equal to
\begin{equation}
\frac{1}{2}\mathrm{Sp}g\left( \mathrm{abs}\left( \Delta ^{-1}\alpha \right)
-I/2\right) =\frac{1}{2}\sum_{j=1}^{2s}g(|\lambda _{j}|-\frac{1}{2}),
\label{entropy}
\end{equation}%
where $g(x)=(x+1)\log (x+1)-x\log x$, Sp denotes trace of the matrices as
distinct from that of operators in $\mathcal{H}$, and $\lambda _{j}$ are the
eigenvalues of $\Delta ^{-1}\alpha $ (see e.g. \cite{H-SSQT}, Sec. 12.3.4),
we have
\begin{eqnarray}
\sup_{\rho :\mathrm{Tr}\rho H\leq E}S(\Phi \lbrack \rho ]) &=&\frac{1}{2}%
\sup_{\beta :\mathrm{Sp}K\epsilon K^{t}\beta \leq E}\mathrm{Sp}g\left(
\mathrm{abs}\left( \Delta _{B}^{-1}\left( K^{t}\beta K+\alpha \right)
\right) -I/2\right)   \notag \\
&=&\frac{1}{2}\max_{\mu :\mathrm{Sp}\epsilon \mu \leq E}\mathrm{Sp}g\left(
\mathrm{abs}\left( \Delta _{B}^{-1}\left( \mu +\alpha \right) \right)
-I/2\right) .  \label{max}
\end{eqnarray}%
Here in the first equality we used the formula (\ref{entropy}) for the
output state with the covariance matrix $K^{t}\beta K+\alpha ,$ and in the
second we denoted $\mu =K^{t}\beta K$ and used the fact that for every $\mu $
such a $\beta $ exists due to the condition $\mathbf{Ran}K^{t}=Z_{B}$. In
the second expression the supremum is attained on some $\mu _{0}$ due to
nondegeneracy of $\epsilon$ (see \cite{H-SSQT}, Sec. 12.5). Denote by $\beta _{0}$ a solution of the
equation $\mu _{0}=K^{t}\beta _{0}K.$

We construct a sequence of suboptimal ensembles as follows. Using the
condition 1 of the Lemma, we let $\rho _{n}$ be a centered Gaussian state in $\mathcal{H}_{A}
$ with the covariance matrices $\gamma _{n}$ and $\rho
_{n}(z)=D(z)\rho _{n}D(z)^{\ast },z\in Z_{A},$ be the family of the
displaced states, where $D(z)$ are the displacement operators obtained by
re-parametrization of the Weyl operators $W(z)$. Define the Gaussian
probability density $p_{n}(z)$ with zero mean and the covariance matrix $%
k_{n}\beta _{0},$ where $k_{n}=1-\mathrm{Sp}\gamma _{n}K\epsilon K^{t}/E>0$
for large enough $n$ by the condition 2$.$ The average state of this
ensemble is centered Gaussian with the covariance matrix $\gamma
_{n}+k_{n}\beta _{0}.$ Taking into account that $S(\rho _{n}(z))=S(\rho
_{n}),$ the $\chi -$quantity of this ensemble is equal to
\begin{equation*}
\chi _{n}=\frac{1}{2}\mathrm{Sp}\,g\left( \mathrm{abs}\left( \Delta
_{B}^{-1}\left( K^{t}\gamma _{n}K+k_{n}K^{t}\beta _{0}K+\alpha \right)
\right) -I/2\right)
\end{equation*}%
\begin{equation*}
-\frac{1}{2}\mathrm{Sp}\,g\left( \mathrm{abs}\left( \Delta _{B}^{-1}\left(
K^{t}\gamma _{n}K+\alpha \right) \right) -I/2\right) .
\end{equation*}%
By the condition 2 this converges to
\begin{equation*}
\frac{1}{2}\mathrm{Sp}\,g\left( \mathrm{abs}\left( \Delta _{B}^{-1}\left(
K^{t}\beta _{0}K+\alpha \right) \right) -I/2\right) -\frac{1}{2}\mathrm{Sp}%
\,g\left( \mathrm{abs}\left( \Delta _{B}^{-1}\alpha \right) -I/2\right) .
\end{equation*}%
By minimality of the noise the second term is entropy of a pure state, equal
to zero, and the first term is just the maximum in (\ref{max}). Thus
\begin{equation*}
C(E)\geq \limsup_{n\rightarrow \infty }\chi _{n}=\sup_{\rho :\mathrm{Tr}\rho
H\leq E}S(\Phi \lbrack \rho ]).\qquad \square
\end{equation*}

\section{One mode}\label{3}

Let $q,p$ be a Bosonic mode, $W(z)=\exp i(xq+yp)$ the corresponding Weyl
operator and $D(z)=\exp i(yq-xp)$ the displacement operator. We give two
examples where the channel describes classical signal with additive Gaussian
(minimal) quantum noise, in the first case the signal being two-dimensional
while in the second -- one-dimensional. As we have seen, a c-q channel can
be described in two equivalent ways: as a mapping $m\rightarrow \rho _{m},$
where $m$ is the classical signal, and as an extended quantum channel
satisfying (\ref{cl-qu}).

1. We first consider the minimal noise c-q channel with two-dimensional real
signal and show the coincidence of the classical entanglement-assisted and
unassisted capacities of this channel under appropriate input constraint, by
using result of Sec. \ref{2}. Such a coincidence is generic for
unconstrained finite-dimensional channels \cite{E}, but in infinite
dimensions, as we will see in the second example, situation is different.
Some sufficient conditions for the equality $C=C_{ea}$ were given in \cite%
{Sh-19}, however they do not apply here.

Let $m=(m_{q},m_{p})\in \mathbf{R}^{2}$ and consider the mapping $%
m\rightarrow \rho _{m}$, where $\rho _{m}$ is the state with the
characteristic function
\begin{equation}
\mathrm{Tr}\rho _{m}W(z)=\exp \left[ i(m_{q}x+m_{p}y)-\frac{\left( N+\frac{1%
}{2}\right) }{2}(x^{2}+y^{2})\right] ,  \label{component}
\end{equation}%
so that%
\begin{equation*}
\rho _{m}=D(m)\rho _{0}D(m)^{\ast }.
\end{equation*}%
The mapping $m\rightarrow \rho _{m}$ can be considered as transmission of
the two-dimensional classical signal $m=(m_{q},m_{p})$ with the additive
quantum Gaussian noise $q,p$ with the average number of quanta $N$. The
minimal noise corresponds to $N=0$.

The classical capacity of this channel with the input constraint
\begin{equation}
\frac{1}{2}\int \left\Vert m\right\Vert ^{2}\,\,p(m)d^{2}m\leq E  \label{3-5}
\end{equation}%
is given by the expression (see e.g. \cite{H-SSQT}, Sec. 12.1.4)
\begin{equation*}
C(E)=g(N+E)-g(N),
\end{equation*}%
with the optimal distribution
\begin{equation}
p(m)=\frac{1}{2\pi E}\,\mbox{exp}\left( -\frac{\left\Vert m\right\Vert ^{2}}{%
2E}\right)   \label{3-9}
\end{equation}
in the ensemble of coherent states $|m\rangle\langle m|$.
In particular, for the minimal noise channel ($N=0$),
\begin{equation}
C(E)=g(E)=S(\bar{\rho}),  \label{cge}
\end{equation}%
where $\bar{\rho}$ is the Gaussian state with
\begin{equation*}
\mathrm{Tr}\bar{\rho}W(z)=\exp \left[ -\frac{\left( E+\frac{1}{2}\right) }{2}%
(x^{2}+y^{2})\right] .
\end{equation*}

Let us now embed this channel into quantum
Gaussian channel $\Phi $ in the spirit of previous Section. Since the input $%
m=(m_{q},m_{p})$ is two-dimensional classical, one has to use two Bosonic
input modes $q_{1},p_{1,},q_{2},p_{2}$ to describe it quantum-mechanically,
so that e.g. $m_{q}=q_{1},m_{p}=q_{2}.$ The environment is one mode $q,p$ in
the Gaussian state $\rho _{0}$ so the output is given by the equations
\begin{eqnarray}
q^{\prime } &=&q+q_{1}=q+m_{q};  \label{eqcha} \\
p^{\prime } &=&p+q_{2}=p+m_{p},  \notag
\end{eqnarray}%
and the channel $\Phi $ parameters are
\begin{equation*}
K=\left[
\begin{array}{cc}
1 & 0 \\
0 & 0 \\
0 & 1 \\
0 & 0%
\end{array}%
\right] ,\quad \alpha =\left( N+\frac{1}{2}\right) I_{2}.
\end{equation*}%
The equations for the environment modes describing the weakly complementary
channel $\tilde{\Phi}^{w}$ are
\begin{eqnarray}
q_{1}^{\prime } &=&q_{1},  \label{eqen} \\
p_{1}^{\prime } &=&p_{1}-p-q_{2}/2,  \notag \\
q_{2}^{\prime } &=&q_{2},  \notag \\
p_{2}^{\prime } &=&p_{2}+q+q_{1}/2.  \notag
\end{eqnarray}%
In fact, the set of equations (\ref{eqcha}), (\ref{eqen}) is the same as for
the quantum channel with additive classical Gaussian noise (see \cite{H-SSQT}%
, Ex. 12.42), but in the latter case the input variables are $q,p$ while in
the former -- $q_{1},p_{1,},q_{2},p_{2}$ (in both cases the output is $%
q^{\prime },p^{\prime }$). If $N=0$ so that $\rho _{0}$ is pure, these
equations describe the complementary channel $\tilde{\Phi}$.

Having realized the c-q channel as a quantum one (i.e. a channel with
quantum input and output), it makes sense to speak of its
entanglement-assisted capacity. Under the same constraint it is given by the
expression
\begin{equation}
C_{ea}(E)=\sup_{\rho _{12}\in \mathfrak{S}_{E}}I(\rho _{12},\Phi ),
\label{cea}
\end{equation}%
where
\begin{equation*}
\mathfrak{S}_{E}=\left\{ \rho _{12}:\mathrm{Tr}\rho _{12}\left( \frac{%
q_{1}^{2}+q_{2}^{2}}{2}\right) \leq E\right\}
\end{equation*}%
corresponds to the constraint (\ref{3-5}). Notice that the constraint
operator $H=\frac{q_{1}^{2}+q_{2}^{2}}{2}$ is unusual in that it is given by
\emph{degenerate} quadratic form in the input variables $%
q_{1},p_{1,},q_{2},p_{2}$. In this case the set $\mathfrak{S}_{E}$ is not
compact, the supremum in (\ref{cea}) is not attained and
to obtain this formula we need to use a result from \cite{H-Sh}.

Now assume the minimal noise $N=0$ and let us show that
\begin{equation}
C_{ea}(E)=C(E)=g(E).  \label{ccea}
\end{equation}%
Proposition 1 of Sec. \ref{2} implies
\begin{equation*}
C_{ea}(E)\leq \sup_{\rho _{12}\in \mathfrak{S}_{E}}S(\Phi \lbrack \rho
_{12}]).
\end{equation*}%
But
\begin{equation*}
\Phi \lbrack \mathfrak{S}_{E}]=\left\{ \bar{\rho}_{p}:p\in \mathcal{P}%
_{E}\right\},
\end{equation*}%
where $\mathcal{P}_{E}$ is defined by (\ref{3-9}), as can be seen from the
equations of the channel (\ref{eqcha}) and the identification of the
probability density $p(m_{q}\,,m_{p})$ with that of observables $q_{1},q_{2}$
in the state $\rho _{12}.$ Invoking (\ref{cge}) gives $\sup_{\rho _{12}\in
\mathfrak{S}_{E}}H(\Phi \lbrack \rho _{12}])=g(E)$ and hence the equality (%
\ref{ccea}). This example is a special case of Proposition 2 in Sec. \ref%
{2}, all the conditions of which are fulfilled with $\mathbf{Ran}K^t=Z_{B}=%
\mathbf{R}^2$ and
\begin{equation*}
\gamma _{n}=\left[
\begin{array}{cccc}
\varepsilon _{n} & 0 & 0 & 0 \\
0 & \frac{1}{4\varepsilon _{n}} & 0 & 0 \\
0 & 0 & \varepsilon _{n} & 0 \\
0 & 0 & 0 & \frac{1}{4\varepsilon _{n}}%
\end{array}
\right] .
\end{equation*}

2. Now we give an example with $C(E)<C_{ea}(E).$ Let $m\in \mathbf{R}$ be a
real one-dimensional signal and the channel is $m\rightarrow \rho _{m}$,
where $\rho _{m}$ is the state with the characteristic function
\begin{equation}
\mathrm{Tr}\rho _{m}W(z)=\exp \left[ imx-\frac{1}{2}(\sigma ^{2}x^{2}+\frac{1%
}{4\sigma ^{2}}y^{2})\right] ,  \label{component1}
\end{equation}%
so that%
\begin{equation*}
\rho _{m}=D(x,0)\rho _{0}D(x,0)^{\ast }.
\end{equation*}%
The mapping $m\rightarrow \rho _{m}$ can be considered as transmission of
the classical signal $m$ with the additive  noise arising from the $q$-component of quantum Gaussian
mode $q,p$ with
the variances $\mathsf{D}q=\sigma ^{2},\mathsf{D}p=\frac{1}{4\sigma ^{2}}$
and zero covariance between $q$ and $p$. The state $\rho _{0}$ is pure
(squeezed vacuum) corresponding to a minimal noise.

The constraint on the input probability distribution $p(m)$ is defined as%
\begin{equation}
\int m^{2}\,\,p(m)dm\leq E,  \label{3-51}
\end{equation}%
where $E$ is a positive constant. As the component $p$ is not affected by
the signal, from information-theoretic point of view this channel is
equivalent to the classical additive Gaussian noise channel $m\rightarrow
m+q,$ and its capacity under the constraint (\ref{3-51}) is given by the
Shannon formula
\begin{equation}
C(E)=\frac{1}{2}\log \left( 1+r\right) ,  \label{gacapa1}
\end{equation}%
where $r=E/\sigma ^{2}$ is the \emph{signal-to-noise ratio}.

A different way to describe this channel is to represent it as a quantum
Gaussian channel $\Phi $. Introducing the input mode $q_{1},p_{1},$ so that $%
m=q_{1},$ with the environment mode $q,p$ in the state $\rho _{0}$, the
output is given by the equations
\begin{eqnarray}
q_{1}^{\prime } &=&q_{1}+q;  \label{eqcha1} \\
p_{1}^{\prime } &=&\quad \quad p,  \notag
\end{eqnarray}%
and the channel $\Phi $ parameters are
\begin{equation*}
K=\left[
\begin{array}{cc}
1 & 0 \\
0 & 0%
\end{array}%
\right] ,\quad \alpha =\left[
\begin{array}{cc}
\sigma ^{2} & 0 \\
0 & \frac{1}{4\sigma ^{2}}%
\end{array}%
\right] .
\end{equation*}%
The equations for the environment mode describing the complementary channel $%
\tilde{\Phi}$ are (see \cite{H-SSQT})
\begin{eqnarray}
q^{\prime } &=&q_{1},  \label{eqen1} \\
p^{\prime } &=&p_{1}-p,  \notag
\end{eqnarray}%
and the set of equations (\ref{eqcha1}), (\ref{eqen1}) describes the
canonical transformation of the composite system = system+environment.

The classical entanglement-assisted capacity of this channel under the same
constraint is given by the expression
\begin{equation}
C_{ea}(E)=\sup_{\rho _{1}\in \mathfrak{S}_{E}^{(1)}}I(\rho _{1},\Phi ),
\label{cea1}
\end{equation}%
where $\mathfrak{S}_{E}^{(1)}=\left\{ \rho _{1}:\mathrm{Tr}\rho
_{1}q_{1}^{2}\leq E\right\} .$ As in the first example, the constraint
operator $q_{1}^{2}$ is given by degenerate quadratic form in the input
variables $q_{1},p_{1}$, the set $\mathfrak{S}_{E}^{(1)}$ is not compact and
the supremum in (\ref{cea}) is not attained.

Let us compute the entanglement-assisted capacity. For this consider the
values of $I(\rho _{A},\Phi )$ for centered Gaussian states $\rho _{A}=\rho
_{1}$ with covariance matrices
\begin{equation*}
\alpha _{1}=\left[
\begin{array}{cc}
E & 0 \\
0 & E_{1}%
\end{array}%
\right] ,
\end{equation*}%
satisfying the uncertainty relation $EE_{1}\geq \frac{1}{4}$ and belonging
to the set $\mathfrak{S}_{E}^{(1)}$ with the equality.

We use the formula (\ref{entropy}) implying
\begin{equation*}
S(\rho _{A})=g\left( \sqrt{EE_{1}}-\frac{1}{2}\right) ,
\end{equation*}%
According to (\ref{eqcha1}), the output state $\rho _{B}=\Phi \lbrack \rho
_{A}]$ has the covariance matrix
\begin{equation*}
\alpha _{B}=\left[
\begin{array}{cc}
E+\sigma ^{2} & 0 \\
0 & \frac{1}{4\sigma ^{2}}%
\end{array}%
\right] ,
\end{equation*}%
with the entropy
\begin{equation*}
S(\rho _{B})=g\left( \sqrt{\frac{E}{4\sigma ^{2}}+\frac{1}{4}}-\frac{1}{2}%
\right) .
\end{equation*}%
Similarly, according to (\ref{eqen1}) the state $\rho _{E}=\tilde{\Phi}[\rho _{A}]$ of
the environment has the covariance matrix
\begin{equation*}
\alpha _{E}=\left[
\begin{array}{cc}
E & 0 \\
0 & E_{1}+\frac{1}{4\sigma ^{2}}%
\end{array}%
\right] ,
\end{equation*}%
with the entropy
\begin{equation*}
S(\rho _{E})=g\left( \sqrt{EE_{1}+\frac{E}{4\sigma ^{2}}}-\frac{1}{2}\right)
.
\end{equation*}%
Summing up,
\begin{eqnarray*}
I(\rho _{A},\Phi ) &=&S(\rho _{A})+S(\rho _{B})-S(\rho _{E}) \\
&=&g\left( \sqrt{\frac{E}{4\sigma ^{2}}+\frac{1}{4}}-\frac{1}{2}\right)
-\delta _{1}(E_{1}),
\end{eqnarray*}%
where
\begin{equation*}
\delta _{1}(E_{1})=g\left( \sqrt{EE_{1}+\frac{E}{4\sigma ^{2}}}-\frac{1}{2}%
\right) -g\left( \sqrt{EE_{1}}-\frac{1}{2}\right)
\end{equation*}%
is a positive function in the range $[\frac{1}{4E},\infty ),$ decreasing
from $g\left( \sqrt{\frac{E}{4\sigma ^{2}}+\frac{1}{4}}-\frac{1}{2}\right) $
to 0 for $E_{1}\rightarrow \infty $ (this follows from the asymptotic $%
g\left( x\right) =\log \left( x/\mathrm{e}\right) +o(1)$). Thus%
\begin{equation*}
C_{ea}(E)\geq g\left( \sqrt{\frac{E}{4\sigma ^{2}}+\frac{1}{4}}-\frac{1}{2}%
\right) .
\end{equation*}

Let us show that in fact there is equality here, by using the concavity of
the quantum mutual information (see \cite{H-SSQT}, Sec. 12.5). For a given input state $\rho $ with finite
second moments consider the state
\begin{equation*}
\tilde{\rho}=\frac{1}{2}\left( \rho +\rho ^{\top }\right) ,
\end{equation*}%
where the transposition $^{\top }$ corresponds to the antiunitary
conjugation $q,p\rightarrow q,-p.$ The state $\tilde{\rho}$ has the same
variances $\mathsf{D}q, \mathsf{D}p$ as $\rho$, and zero covariance between $q$ and $p$.
The channel (\ref{eqcha1}) is covariant with respect to the transposition;
by the aforementioned concavity, $I(\tilde{\rho},\Phi )\geq I(\rho ,\Phi ),$
moreover, $I(\tilde{\rho}_{G},\Phi )\geq I(\tilde{\rho},\Phi ),$ where $%
\tilde{\rho}_{G}$ is the Gaussian state with the same first and second
moments as $\tilde{\rho}.$ Thus
\begin{eqnarray*}
C_{ea}(E) &=&g\left( \sqrt{\frac{E}{4\sigma ^{2}}+\frac{1}{4}}-\frac{1}{2}%
\right) =g\left( \frac{\sqrt{1+r}-1}{2}\right) \\
&=&\frac{\sqrt{1+r}+1}{2}\log \frac{\sqrt{1+r}+1}{2}-\frac{\sqrt{1+r}-1}{2}%
\log \frac{\sqrt{1+r}-1}{2},
\end{eqnarray*}
where $r=E/\sigma ^{2}$ is signal-to-noise ratio. Comparing this with (\ref%
{gacapa1}), one has $C_{ea}(E)>C(E)$ for $E>0$ (see Appendix), with the
entanglement-assistance gain $C_{ea}(E)/C(E)\sim -\frac{1}{2}\log r,$ as $%
r\rightarrow 0$ and $C_{ea}(E)/C(E)\rightarrow 1,$ as $r\rightarrow \infty $
(see Figures).

As it is to be expected, Proposition 2 is not applicable, as $\mathrm{rank}%
K^t=1<\dim Z_{B}$ here, while
\begin{equation*}
\gamma _{n}=\left[
\begin{array}{cc}
\varepsilon _{n} & 0 \\
0 & \frac{1}{4\varepsilon _{n}}%
\end{array}%
\right]
\end{equation*}%
still satisfies the conditions 1, 2 of the Lemma.

\section{Appendix}

1. Consider the channel (\ref{eqcha}). It is instructive to compare its unassisted classical capacity $C(E)$
given by (\ref{ccea}) with the values of $I(\rho _{12},\Phi )$ for centered Gaussian states $\rho
_{12}=\rho _{A}$ with the covariance matrices
\begin{equation*}
\alpha _{12}=\left[
\begin{array}{cccc}
E & 0 & 0 & 0 \\
0 & E_{1} & 0 & 0 \\
0 & 0 & E & 0 \\
0 & 0 & 0 & E_{1}%
\end{array}%
\right] ,
\end{equation*}%
satisfying the uncertainty relation $EE_{1}\geq \frac{1}{4}$ and belonging
to the set $\mathfrak{S}_{E}$ with the equality.

We then find
\begin{equation*}
S(\rho _{12})=2g\left( \sqrt{EE_{1}}-\frac{1}{2}\right) .
\end{equation*}%
According to (\ref{eqcha}), $\rho _{B}=\Phi \lbrack \rho _{A}]$ has the
covariance matrix
\begin{equation*}
\alpha _{B}=\left[
\begin{array}{cc}
E+\frac{1}{2} & 0 \\
0 & E+\frac{1}{2}%
\end{array}%
\right] ,
\end{equation*}%
with the entropy $g(E),$ and according to (\ref{eqen}) the state $\rho _{E}$
of the environment has the covariance matrix
\begin{equation*}
\alpha _{E}=\left[
\begin{array}{cccc}
E & 0 & 0 & E/2 \\
0 & \tilde{E}_{1} & -E/2 & 0 \\
0 & -E/2 & E & 0 \\
E/2 & 0 & 0 & \tilde{E}_{1}%
\end{array}%
\right] ,
\end{equation*}%
where $\tilde{E}_{1}=E_{1}+\frac{1}{2}+\frac{E}{4}.$ The eigenvalues of $%
\Delta _{E}^{-1}\alpha _{E}$ are $\sqrt{E}\left( \sqrt{\tilde{E}_{1}}\pm
\frac{1}{2}\sqrt{E}\right) $ and have multiplicity 2. Thus
\begin{equation*}
S(\rho _{E})=S(\tilde{\Phi}[\rho _{12}])=g\left( \sqrt{E}\left( \sqrt{\tilde{%
E}_{1}}+\frac{1}{2}\sqrt{E}\right) -\frac{1}{2}\right)
\end{equation*}%
\begin{equation*}
+g\left( \sqrt{E}\left( \sqrt{\tilde{E}_{1}}-\frac{1}{2}\sqrt{E}\right) -%
\frac{1}{2}\right) .
\end{equation*}%
Summing up,
\begin{equation*}
I(\rho _{12},\Phi )=g(E)-\delta (E_{1}),
\end{equation*}%
where
\begin{eqnarray*}
\delta (E_{1}) &=&g\left( \sqrt{E}\left( \sqrt{\tilde{E}_{1}}+\frac{1}{2}%
\sqrt{E}\right) -\frac{1}{2}\right) +g\left( \sqrt{E}\left( \sqrt{\tilde{E}%
_{1}}-\frac{1}{2}\sqrt{E}\right) -\frac{1}{2}\right) \\
&&-2g\left( \sqrt{EE_{1}}-\frac{1}{2}\right)
\end{eqnarray*}%
is a positive function in the range $[\frac{1}{4E},\infty ),$ varying from $%
g(E)$ to 0. Hence the value (\ref{ccea}) is attained only asymptotically for
the input states $\rho _{12}$ with momentum variance $E_{1}\rightarrow
\infty .$

2. Introducing the new variable $x=\sqrt{1+r}\geq 1,$ we have
\begin{equation*}
C(E)=\log x\equiv f_{1}(x),\quad C_{ea}(E)=\frac{x+1}{2}\log \frac{x+1}{2}-%
\frac{x-1}{2}\log \frac{x-1}{2}\equiv f_{2}(x).
\end{equation*}%
Then $f_{1}(1)=f_{2}(1),f_{1}^{\prime }(\infty )=f_{2}^{\prime }(\infty )$
and $f_{1}^{\prime \prime }(x)>f_{2}^{\prime \prime }(x).$ It follows $%
f_{1}(x)<f_{2}(x),x>1.$ \bigskip

\textbf{Acknowledgments.} This work was partly supported by RFBR grant N
12-01-00319-a, Fundamental Research Programs of RAS and by the Cariplo
Fellowship under the auspices of the Landau Network - Centro Volta. The
author is grateful to G. M. D'Ariano for the hospitality at the QUIT group
of the University of Pavia, and to A. Barchielli, L. Maccone, P.
Perinotti, M.F. Sacchi and M.E. Shirokov for stimulating discussions.
Special thanks are due to L. Maccone for the help with Latex graphics.

\newpage
\begin{figure}[tbp]
\centering  \includegraphics[width=6cm]{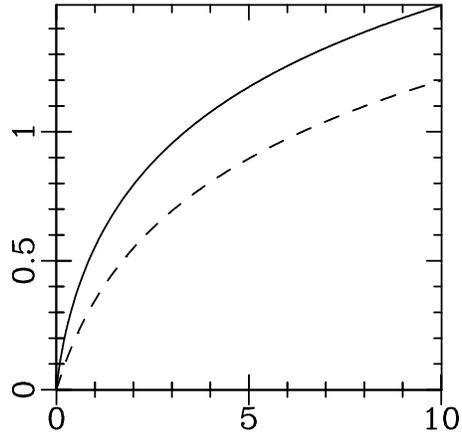} 
\caption{\emph{Ex.2: The classical capacities (nats) as functions of
signal-to-noise ratio $r$: $C_{ea}(E)$ -- solid line, $C(E)$ -- dashed line.}}
\end{figure}
\begin{figure}[tbp]
\centering  \includegraphics[width=6cm]{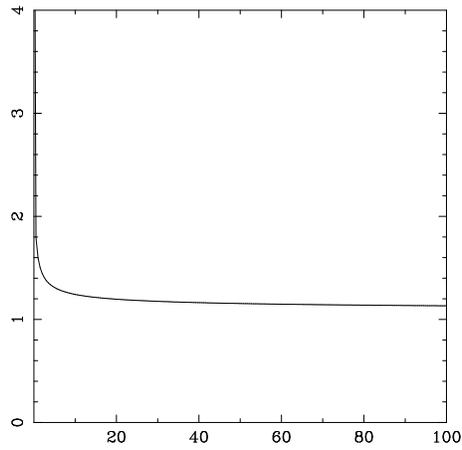} 
\caption{\emph{Ex.2: The gain of entanglement assistance $C_{ea}(E)/C(E)$
\newline as function of signal-to-noise ratio.}}
\end{figure}


\begin{thebibliography}{9}
\bibitem{BSST} C.H. Bennett, P.W. Shor, J.A.Smolin, A.V.Thapliyal,
Entanglement-assisted classical capacity of noisy quantum channel, Phys.
Rev. Lett. \textbf{83}, 3081-3084, (1999); arXiv:quant-ph/9904023.

\bibitem{E} C.H. Bennett, P.W. Shor, J.A.Smolin, A.V.Thapliyal,
Entanglement-assisted capacity and the reverse Shannon theorem, IEEE Trans.
Inform. Theory \textbf{48} 2637-2655, (2002); arXiv:quant-ph/0106052.

\bibitem{cegh1} F. Caruso, J. Eisert, A.S.Holevo, V. Giovannetti, The
optimal unitary dilation for bosonic Gaussian channels, Phys. Rev. A \textbf{%
84}, 022306 (2011).

\bibitem{H-SSQT} A.S. Holevo, Quantum systems, channels, information. A
mathematical introduction, Berlin/Boston, DeGruyter, 2012.

\bibitem{H1} A.S. Holevo, Entropy gain and the Choi-Jamiolkowski
correspondence for infinite-dimensional quantum evolutions, Theoretical and
Mathematical Physics, \textbf{166}:1 (2011), 123-138; arXiv:1003.5765 [quant-ph].

\bibitem{H2} A.S. Holevo, Information capacity of quantum observable, Probl.
Inform. Transmission \textbf{48}:1 (2012); arXiv:1103.2615 [quant-ph].

\bibitem{H3} A.S. Holevo, On extreme Bosonic linear channels,
arXiv:1111.3552 [quant-ph].

\bibitem{H-Sh} A.S. Holevo, M.E. Shirokov, On the entanglement-assisted
classical capacity of infinite-dimensional quantum channels, arXiv:1210.6926.

\bibitem{Sh-19} M.E. Shirokov, Conditions for equality between
entanglement-assisted and unassisted classical capacities of a quantum
channel, Probl. Inform. Transm. \textbf{48}:2 85-101 (2012).
\end{thebibliography}
\end{document}